\documentclass[twocolumn]{article}
\usepackage{times,bm}
\usepackage{amsmath}
\usepackage[affil-it,]{authblk}
\usepackage[pagebackref=false,colorlinks=true,linkcolor=acsblue,citecolor=olive,urlcolor=acsblue]{hyperref}

\usepackage{geometry}

\usepackage[switch*, displaymath,pagewise, mathlines]{lineno}
\usepackage{graphicx}
\usepackage{cite}
\usepackage{amsmath}
\usepackage{amsfonts}
\usepackage{amssymb}
\usepackage{amsmath} 
\usepackage{slashed}
\usepackage{subfloat}
\usepackage{slashed}
\usepackage{color}
\usepackage{float}
\usepackage{multirow,multicol}
\usepackage{tabulary}
\usepackage[flushleft]{threeparttable}
\usepackage{pgfplots,subfigure}
\usepackage{xcolor}
\numberwithin{equation}{section}
\usepackage{tikz}
\usepackage{ulem}
\usepackage{hyperref}
\usepackage{nameref}
\usepackage{comment}

\usepgflibrary{arrows}
\usetikzlibrary{shapes.callouts}
\tikzset{
	level/.style   = { thick, },
	connect/.style = { dotted, red   },
	notice/.style  = { draw, rectangle callout, callout relative pointer={#1} },
	label/.style   = { text width=1cm }
}

\definecolor{acsblue}{RGB}{17,76,139}
\definecolor{shadecolor}{RGB}{255,241,204}
\usepackage{letltxmacro}
\makeatletter
\let\oldr@@t\r@@t
\def\r@@t#1#2{%
	\setbox0=\hbox{$\oldr@@t#1{#2\,}$}\dimen0=\ht0
	\advance\dimen0-0.2\ht0
	\setbox2=\hbox{\vrule height\ht0 depth -\dimen0}%
	{\box0\lower0.4pt\box2}}
\LetLtxMacro{\oldsqrt}{\sqrt}
\renewcommand*{\sqrt}[2][\ ]{\oldsqrt[#1]{#2}}
\makeatother
\usepackage{environ}
\usepackage{breqn}

\begin{document}

\newcommand{{\ri}}{{\rm{i}}}
\newcommand{{\Psibar}}{{\bar{\Psi}}}
\newcommand*\var{\mathit}

\newcommand{\myspace}{\,}
%
\fontsize{7.6}{8.6}\selectfont

\title{\mdseries{Geometric and wave optics in a BTZ optical metric-based wormhole}}

\author{ \textit {\mdseries{Semra Gurtas Dogan}}$^{\ 1}$\footnote{\textit{ E-mail: semragurtasdogan@hakkari.edu.tr (Corr. Author)} }~,~ \textit {\mdseries{Abdullah Guvendi}}$^{\ 2}$\footnote{\textit{E-mail: abdullah.guvendi@erzurum.edu.tr } }~,~ \textit {\mdseries{Omar Mustafa}}$^{\ 3}$\footnote{\textit{ E-mail: omar.mustafa@emu.edu.tr} }  \\
	\small \textit {$^{\ 1}$\footnotesize Department of Medical Imaging Techniques, Hakkari University, 30000, Hakkari, Türkiye}\\
	\small \textit {$^{\ 2}$\footnotesize Department of Basic Sciences, Erzurum Technical University, 25050, Erzurum, Türkiye}\\
	\small \textit {$^{\ 3}$\footnotesize Department of Physics, Eastern Mediterranean University, 99628, G. Magusa, north Cyprus, Mersin 10 - Türkiye}}
\date{}
\maketitle

\begin{abstract}
We investigate the geometric and wave optical properties of a $(2+1)$-dimensional ultra-static spacetime conformally related to the static BTZ black hole, characterized by constant negative Gaussian curvature. The associated optical metric defines a hyperbolic wormhole geometry, wherein null geodesics experience a P\"oschl--Teller-type repulsive effective potential that suppresses circular photon orbits and directs all trajectories toward the optical origin. In the wave regime, we reformulate the Helmholtz equation into a Schr\"odinger-like form, revealing a spatially localized effective potential that encodes curvature and angular momentum effects. The resulting refractive index $n(\rho,\omega)$ is both spatially and spectrally dispersive, leading to a position-dependent critical frequency $\omega_c(\rho)$ that delineates the boundary between propagating and evanescent modes. At high frequencies, the medium becomes asymptotically transparent, while for $\omega < \omega_c(\rho)$, waves undergo exponential attenuation. These results demonstrate intrinsic curvature-induced spectral filtering and provide a geometrically tunable framework for analog gravity systems and graphene-based photonic platforms.
\end{abstract}

\begin{small}
\begin{center}
\textit{\fontsize{7.7}{8.7}\selectfont Keywords: Optical wormholes; BTZ black hole; Analog gravity; Geometric optics; Wave optics; Curvature-induced refractive index }	
\end{center}
\end{small}



\section{\mdseries{Introduction}}\label{sec1}

The BTZ black hole, introduced by Bañados, Teitelboim, and Zanelli, provides an exact solution to Einstein’s field equations in (2+1)-dimensional spacetime with a negative cosmological constant, \(\Lambda = -\frac{1}{l^2}\), where \(l\) is the AdS\(_3\) radius \cite{1}. Defined by its mass parameter \(M\) and cosmological constant, the static BTZ black hole carries neither angular momentum nor electric charge, resulting in a static spacetime that is asymptotically anti-de Sitter with a well-defined horizon structure \cite{2}. Its metric, in standard coordinates \((t, r, \phi)\), is given by \cite{1,2}:
\[
ds^2 = -\left(-M + \frac{r^2}{l^2}\right) dt^2 + \left(-M + \frac{r^2}{l^2}\right)^{-1} dr^2 + r^2 d\phi^2,
\]
where \(t\), \(r\), and \(\phi\) denote the time, radial, and angular coordinates, respectively. The function \(-M + \frac{r^2}{l^2}\) governs the causal structure of the spacetime, with the event horizon located at \(r_+ = l\sqrt{M}\) for \(M > 0\). In the absence of rotation, this geometry lacks inner horizons and ergoregions, simplifying its structure relative to the rotating BTZ solution. Despite this simplicity, the static BTZ black hole preserves fundamental features such as event horizon properties, thermodynamics, and causal behavior, establishing it as a valuable model for studying gravity in lower dimensions \cite{2}. Its analytical tractability has facilitated detailed investigations into black hole thermodynamics \cite{2,3}, quantum gravity \cite{3}, holographic superconductivity \cite{4}, and quantum critical phenomena within the AdS\(_3\)/CFT\(_2\) framework \cite{5}.

\vspace{0.10cm}
\setlength{\parindent}{0pt}

Remarkably, the BTZ black hole has also found applications in condensed matter physics, particularly in graphene systems~\cite{6}. By mapping its conformal equivalent onto the Beltrami pseudosphere, connections emerge between (2+1)-dimensional gravity and the curved geometry of monolayer graphene~\cite{6,7,8}. Notably, the optical metric of the static BTZ black hole corresponds to a curved surface with constant negative Gaussian curvature, closely resembling the geometry of monolayer graphene sheets. Recent studies have identified this (2+1)-dimensional spacetime as an optical or hyperbolic wormhole, characterized by constant negative Gaussian curvature~\cite{9,10,11}. These investigations demonstrate that the tunneling times of charge carriers and photons can be controlled by adjusting the curvature radius along the radial coordinate of the optical wormhole~\cite{9,10,11}, implying that the optical response of such surfaces can, in principle, be tuned to meet specific requirements. Accordingly, identifying the effective background potentials, arising from the wormhole curvature, that govern ray and wave propagation is critical for advancing photonic and nanoscale technologies. The ability to tailor graphene-based materials through curvature, twisting, or rolling enhances their versatility and potential for applications such as ultrafast photodetectors~\cite{12}.

\vspace{0.10cm}
\setlength{\parindent}{0pt}

Separately, the study of geodesics and wave propagation in curved spaces offers a robust framework for analyzing optical phenomena driven by effective background potentials \cite{Rop1,Rop2}. In general relativity, spacetime curvature induced by mass-energy alters electromagnetic wave propagation, influencing wave optics \cite{R1}, gravitational lensing \cite{R2}, scattering processes \cite{R3}, and photon ring formation \cite{R4,R5,guvendi-npb}. Experimental investigation of these effects remains challenging due to their weak strength, requiring astronomical-scale observations. To address this, analog models have been developed to reproduce relativistic effects in laboratory environments \cite{semra-2025,R6,R7}. Examples include observing spontaneous Hawking radiation in Bose-Einstein condensates \cite{R8}, simulating Schwarzschild precession using gradient-index lenses \cite{R9}, and emulating gravitational lensing with microstructured optical waveguides \cite{R10}. Another approach involves dimensional reduction of curved spacetimes embedded in flat three-dimensional space, facilitating detailed study of electromagnetic wave interactions with curvature \cite{R11,R12}. This method, introduced by Batz and Peschel \cite{R13}, has inspired theoretical and experimental advances across various optical systems \cite{12,R6,R14,R15}. Applications now extend to surface plasmon polaritons \cite{R16} and quantum particle dynamics \cite{R17}. While Batz and Peschel derived a nonlinear Schrödinger equation for only waves on surfaces of revolution with constant Gaussian curvature \cite{R13}, their analysis applies mainly to longitudinal propagation due to the axisymmetric geometry. 

\vspace{0.10cm}
\setlength{\parindent}{0pt}

Here, we perform a thorough analysis of null geodesics and wave dynamics in an optical wormhole realized by the optical metric of the static BTZ black hole, embodying hyperbolic wormhole geometry with constant negative Gaussian curvature. Section \ref{sec:2} revisits the underlying ultra-static spacetime formalism. Section \ref{sec:3} details the characterization of light trajectories. Section \ref{sec:4} addresses wave dynamics through an in-depth study of the Helmholtz equation, emphasizing the influence of effective background potentials on the optical response. Final conclusions and broader physical interpretations are presented in Section \ref{sec:conc}.

\begin{figure}[ht]
\centering
\includegraphics[scale=0.55]{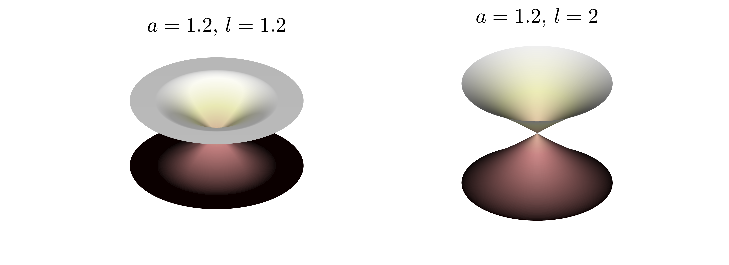}
\caption{ \fontsize{7.7}{8.7}\selectfont Visualization of the wormhole geometry for varying parameter values. The coordinates are defined by \(x(\rho, \phi) = l \sinh\left(\frac{a}{l} \rho\right) \cos(\phi)\), \(y(\rho, \phi) = l \sinh\left(\frac{a}{l} \rho\right) \sin(\phi)\), and \(z(\rho) = \pm \int d\rho \sqrt{1 - \left(\eta^{'}(\rho)\right)^2}\), where the prime denotes the derivative with respect to \(\rho\).}
\label{fig:wormhole}
\end{figure}

\section{\mdseries{Revisiting the BTZ optical metric}} \label{sec:2}

In this section, we revisit the optical metric of the static BTZ black hole, as introduced in~\cite{7}, and demonstrate that this metric describes an optical wormhole, referred to as a hyperbolic wormhole, a name derived from its radial shape function. Recent studies~\cite{6,7} propose that a graphene structure shaped as a surface of revolution with constant negative curvature, such as the Beltrami trumpet, may exhibit phenomena analogous to those observed in quantum field theory on curved spacetimes. Of particular interest is the potential connection to BTZ black hole metrics \cite{1}. These solutions have been extensively studied due to their significance in string theory, particularly in the context of black holes and their quantum microstates. The metric of the static BTZ black hole can be rewritten as follows~\cite{7}:
\begin{equation}
    ds^2_{\text{BTZ}} = -\Delta dt^2 + \frac{dr^2}{\Delta} + r^2  d\phi^2,\quad \Delta(r) = \frac{r^2}{l^2} - M.\label{BTZ}
\end{equation}
It is well established (refer to \cite{Op3} for a comprehensive review) that low-energy electronic excitations on a curved graphene sheet \(\Sigma\) obey the massless Dirac equation on \(\mathbb{R} \times \Sigma\) with respect to the induced metric:
\begin{equation}
    ds^2 = -dt^2 + h_{ij} dx^i dx^j, \quad i = 1, 2,\label{USM}
\end{equation}
where, \( h_{ij} \) denotes the metric induced on the surface \( \Sigma \). The scenario examined in \cite{6} involves a surface of revolution embedded in the Euclidean three-space \( \mathbb{E}^3 \), with coordinates \( (x(x^i), y(x^i), z(x^i)) \). The metric given in \eqref{USM} represents an ultra-static metric, characterized as both static and invariant under time translations and time reversal. The general form of a static metric is~\cite{7}:
\begin{equation}
    ds^2 = -V dt^2 + g_{ij} dx^i dx^j.
\end{equation}
For static black holes, we have $V > 0$ outside the horizon. A metric is termed ultra-static if $V = 1$, implying the absence of gravitational redshift and that gravitational forces do not act on a particle at rest in such a spacetime. Since graphene does not exhibit an apparent source of redshifting, the assumption in \cite{6} that $g_{tt} = -1$ is physically well-founded. It is important to emphasize that a black hole metric cannot be ultra-static. However, the massless Dirac equation, $\gamma^\mu \nabla_\mu \Psi=0$, remains invariant under conformal rescalings \cite{7}. Specifically, in $D$ spacetime dimensions, if 
\begin{equation}
    \tilde{g}_{\mu\nu} = \Omega^2 g_{\mu\nu}, \quad \tilde{\Psi} = \frac{1}{\Omega^{(D-1)/2}} \Psi,
\end{equation}
then the equation transforms into:
\begin{equation}
    \gamma^\mu \nabla_\mu \Psi = \frac{1}{\Omega^{(D+1)/2}} \tilde{\gamma}^\mu \tilde{\nabla}_\mu \Psi.
\end{equation}
Furthermore, any static metric is locally conformally ultra-static~\cite{7}:  
\begin{equation}  
    -V dt^2 + g_{ij} dx^i dx^j = V \left\{ -dt^2 + h_{ij} dx^i dx^j \right\},  
\end{equation}  
where \( h_{ij} = \frac{1}{V} g_{ij} \) is referred to as the optical metric \cite{Op4}. This observation implies that, at least in classical contexts, the effects of horizons on massless fermions can be simulated using ultra-static metrics for graphene, as proposed in \cite{6}. However, in quantum field theory, the vacuum state is not necessarily invariant under conformal transformations, making the simulation of Unruh or Hawking radiation effects non-trivial. It is also important to note that not all time-independent (stationary) metrics are static. For example, static black holes in \( 2+1 \) dimensions, including static BTZ black holes, are stationary. Their metrics generally take the form~\cite{7}:  
\begin{equation}  
    ds^2 = -V\,dt^2 + g_{ij} dx^i dx^j.  
\end{equation}
Locally, any stationary metric can be transformed into one of two forms via conformal rescaling \cite{Op4}. The first form is expressed as:
\begin{equation}
    ds^2_R = -dt^2 + a_{ij} dx^i dx^j,
\end{equation}
where $a_{ij} = V^{-1} g_{ij}$. Alternatively, by applying a different conformal rescaling and completing the square, the corresponding form is:
\begin{equation}
    ds^2 = -dt^2 + h_{ij} dx^i dx^j,
\end{equation}
where $h_{ij}$ represents the optical metric \cite{Op4}. The static BTZ black hole metric, given in \eqref{BTZ}, is already in the optical form, apart from a conformal factor $\Delta$ \cite{7}. From this point, we focus on axisymmetric metrics that can be isometrically embedded into Euclidean space $\mathbb{E}^3$ as surfaces of revolution. These metrics are described by \cite{7}:
\begin{equation}
    h_{ij} dx^i dx^j = d\rho^2 + C^2(\rho) d\phi^2, \quad 0 \leq \phi < 2\pi,
\end{equation}
where $C^2(\rho) = x^2 + y^2 = R^2$ and $d\rho^2 = dR^2 + dz^2$. The Gaussian curvature $(K)$ \cite{Op5} is given by:
\begin{equation}
    K = -\frac{\ddot{C}(\rho)}{C(\rho)},
\end{equation}
where the dot denotes differentiation with respect to the variable $\rho$. The Beltrami trumpet, as introduced in \cite{6}, serves as an example of such a surface \cite{7}:
\begin{equation}
C(\rho) = a \exp\left(-\frac{\rho}{a}\right), \quad \rho \geq 0 \implies K = -\frac{1}{a^2}.
\end{equation}
For the static BTZ metric \eqref{BTZ}, where \(0 \leq \phi < 2\pi\), the corresponding optical metric is expressed as:
\begin{equation}
ds^2_o = -dt^2 + \frac{dr^2}{\Omega^2} + \frac{r^2}{\Omega^2} d\phi^2, \quad \Omega^2 = \frac{r^2}{l^2} - M.\label{Omega}
\end{equation}
We now proceed to embed the exact static BTZ optical geometry into \(E^3\), beginning with the static BTZ optical geometry in equation \eqref{Omega}. The metric is then expressed as follows:
\begin{equation}
ds_o^2 = -dt^2 + \frac{l^4}{(r^2 - a^2)^2} dr^2 + \frac{l^2}{r^2 (r^2 - a^2)} d\phi^2, \label{26}
\end{equation}
where \(a = l \sqrt{M}\). Defining \(C^2 = \frac{l^2 r^2}{r^2 - a^2}\), we can rewrite the following relations (for more details see \cite{7}):
\begin{equation}
\begin{split}
&dz^2 + dC^2 = \frac{l^4}{(r^2 - a^2)^2} dr^2 \implies dz^2 = l^2\frac{(r^2-a^2(l^2+a^2))}{(r^2-a^2)^3}dr^2. \label{28}
\end{split}
\end{equation}
This yields
\begin{equation}
\left( \frac{dz}{dC} \right)^2 = \frac{1+\frac{l^2}{a^2}-\frac{C^2}{l^2}}{C^2-l^2}. \label{29}
\end{equation}
Equation \eqref{29} implies that the embedding process must terminate when \(C = \sqrt{l^2 + a^2}\), which corresponds to a radius beyond the horizon where \(C^2 \to \infty\). The radial optical distance, \(\rho\), is then defined by \(d\rho = \frac{l^2}{r^2 - a^2} dr\), and implies the following relationships: \(r = a\,\coth \left( \frac{a}{l} \rho \right)\), and \( C = l \cosh \left( \frac{a}{l} \rho \right)\) \cite{7}. Thus, the optical metric takes the form \cite{7,9,10,11}:
\begin{equation}
ds_o^2 = -dt^2 + d\rho^2 + l^2 \cosh^2 \left( \frac{a}{l} \rho \right) d\phi^2, \label{32}
\end{equation}
while the BTZ metric itself is given by \cite{7}:
\begin{equation}
\begin{split}
&ds_{BTZ}^2 = \mathcal{Q}(\rho)\left[ -dt^2 + d\rho^2 + l^2 \cosh^2 \left( \frac{a}{l} \rho \right) d\phi^2 \right], \\
&\mathcal{Q}(\rho)=a^2 \sinh^2 \left( \frac{a}{l} \rho \right). \label{33}
\end{split}
\end{equation}
It is worth noting that the Gaussian curvature of the spatial part of the BTZ optical metric is constant and negative \cite{9,10,11}. By interpreting \(l\) as the radius of the wormhole at the midpoint (\(\rho = 0\)) between two sheets, and \(l/a\) as the radius of curvature along \(\hat{\rho}\) of the graphene wormhole connecting the two monolayer sheets, the metric in equation \eqref{32} characterizes a hyperbolic wormhole with constant negative Gaussian curvature \cite{9,10,11}. The geometric structure of this wormhole can be seen in Figure \ref{fig:wormhole}.

\section{\mdseries{Ray optics}} \label{sec:3}

\begin{figure}[ht]
\centering
\includegraphics[scale=0.55]{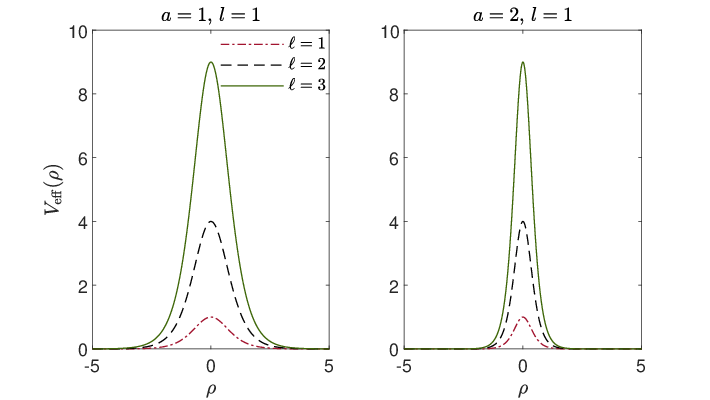}
\caption{ \fontsize{7.7}{8.7}\selectfont Effective potentials $V_{\text{eff}}(\rho)$ for different  $\ell = 1,2,3$ plotted against the radial coordinate $\rho\in[-5,5]$. The two subplots compare potentials for wormhole parameters $a=1, l=1$ (left) and $a=2, l=1$ (right). Increasing $\ell$ raises the potential magnitude, while larger $a$ sharpens the potential barrier around $\rho=0$.}
\label{fig:1}
\end{figure}

In this section, we examine the light propagation in the geometric background described by the metric given in \eqref{32}. We begin by studying the geodesic equation to determine the light ray trajectories. To find the geodesics, we follow the approach outlined in Refs.~\cite{Rop1, Rop2}, where both null and timelike geodesics are determined. We take the Lagrangian as:
\begin{equation}
\begin{split}
&\mathcal{L} = g_{\mu \nu} \frac{dx^{\mu}}{d\lambda} \frac{dx^{\nu}}{d\lambda}, \quad g_{\mu \nu} = \textrm{diag}(-1,\, 1,\, \eta(\rho)^2), \\
&(\mu, \nu = t, \rho, \phi), \quad \eta(\rho) = l \, \cosh \left( \frac{a}{l} \rho \right),
\end{split}
\end{equation}
where $\lambda$ is the affine parameter of the curve, and the geodesics are obtained as solutions of the Euler-Lagrange equation:
\begin{equation}
\frac{\partial \mathcal{L}}{\partial x^{\mu}} - \frac{d}{d\lambda} \left( \frac{\partial \mathcal{L}}{\partial \dot{x}^{\mu}} \right) = 0.
\end{equation}
Now, set $\mathcal{L} = \tilde{\kappa}$ by taking the speed of light ($c$) in vacuum as unity, $c = 1$. It should be noted that the choice $\tilde{\kappa} = 0$ corresponds to lightlike geodesics, while $\tilde{\kappa} = -1$ yields timelike geodesics. Using the line element in Eq.~\eqref{32}, the Lagrangian becomes:
\begin{equation}
\mathcal{L} = -\dot{t}^2 + \dot{\rho}^2 + \eta^2 \dot{\phi}^2,
\end{equation}
where the dot indicates a derivative with respect to the affine parameter \(\lambda\). Since \(\mathcal{L}\) does not explicitly depend on \(t\) and \(\phi\), the derivatives \(\frac{\partial \mathcal{L}}{\partial \dot{t}}\) and \(\frac{\partial \mathcal{L}}{\partial \dot{\phi}}\) are constants of motion, corresponding to the following conserved quantities, respectively:
\begin{equation}
\mathcal{E} = \dot{t}, \quad \ell = \eta^2 \dot{\phi}\,.
\end{equation}
The conserved quantities in question represent the particle's energy and angular momentum, respectively, and are essential for comprehending the underlying nature of geodesic motion. In terms of $\mathcal{E}$ and $\ell$, we can rewrite the Lagrangian as follows:
\begin{equation}
\mathcal{L} = -\mathcal{E}^2 + \dot{\rho}^2 + \frac{\ell^2}{\eta(\rho)^2}.
\end{equation}
This formulation of the Lagrangian provides a more straightforward interpretation of the motion's dynamics, with energy $\mathcal{E}$ and angular momentum $\ell$ acting as constraints. The radial motion is influenced by the potential term $ \ell^2/\eta(\rho)^2$, which is a function of $\eta(\rho)$, and this term contributes to the effective potential encountered by the ray. For null geodesics ($\tilde{\kappa}=0$), we obtain
\begin{equation}
\left(\frac{d\rho}{d\lambda}\right)^2 = \mathcal{E}^2 - \frac{\ell^2}{\eta(\rho)^2}.
\end{equation}
This equation can be interpreted as one-dimensional motion of a particle (with energy $\mathcal{E}$) under the influence of an effective potential given by (see also \cite{Rop2}):
\begin{equation}
V_{\text{eff}}(\rho) = \frac{\ell^2}{l^2\, \cosh^2\left( \frac{a}{l} \rho \right)}.\label{eff-pot}
\end{equation}
This effective potential is a classic example of a Pöschl-Teller potential, widely studied for its exact solvability and relevance in quantum systems. This potential exhibits a finite, localized peak centered at \(\rho = 0\), where it attains its maximum value of \(V_{\text{eff}}(0) = \frac{\ell^2}{l^2}\). The \(\cosh^2\left(\frac{a}{l} \rho\right)\) term governs the spatial decay, ensuring that the potential decreases symmetrically and rapidly toward zero as \(|\rho| \to \infty\). This asymptotic behavior reflects the diminishing influence of the potential at large distances, making it effective only within a finite region around the throat. The parameter \(\ell\) determines the strength of the peak, while \(l\) and \(a\) influence its width and the decay rate. Physically, this potential describes a repulsive interaction that prevents particles from occupying the central region, with no attractive well or stable photon orbits. Its resemblance to the hyperbolic secant-squared function underscores its utility in modeling systems with localized barriers, such as scattering processes or field-theoretic solitonic structures. Additionally, our results imply that radial motion occurs only when \(\frac{d\rho}{d\lambda} > 0\), as the condition \(\mathcal{E}^2 < \frac{\ell^2}{\eta(\rho)^2}\) is prohibited for radial motion. The behavior of the effective potential \(V_{\text{eff}}(\rho)\) is illustrated in Figure \ref{fig:1} for \( \ell = 1 \), \( \ell = 2 \), and \( \ell = 3 \). As depicted, these values of \( \ell \) do not exhibit any stable equilibrium points. Furthermore, the figure implies a significant reduction in the tunneling time of quantum fields as the radius of curvature (\( \ell/a \)) of the optical wormhole connecting two monolayer sheets decreases. This observation aligns with the findings reported in Ref. \cite{9} for Weyl pairs in the hyperbolic (optical) wormhole background. For further details, refer to \cite{9} and \cite{10}. A similar effect has also been observed for photons (massless vector bosons) in such static optical wormholes, as discussed in \cite{11}.

\vspace{0.10cm}
\setlength{\parindent}{0pt}

\begin{figure}[ht]
\centering
\includegraphics[scale=0.50]{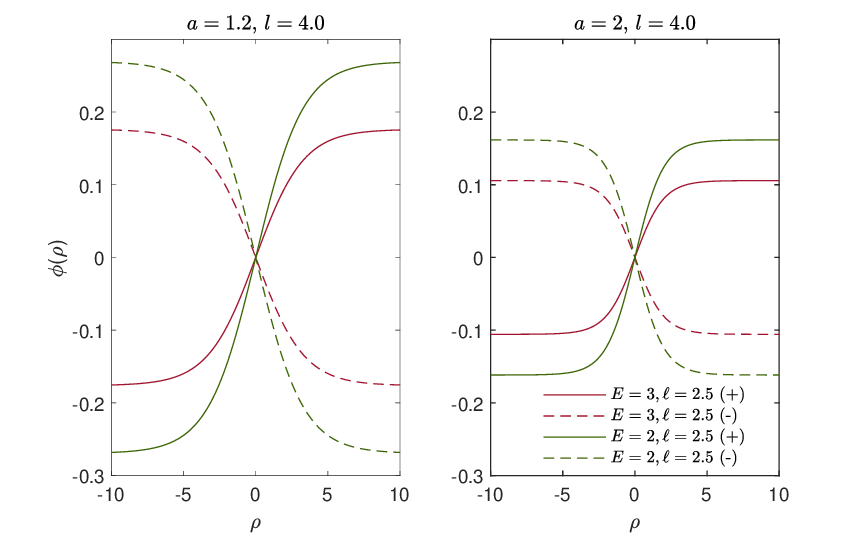}
\caption{ \fontsize{7.7}{8.7}\selectfont Angular trajectories $\phi(\rho)$ of null geodesics around the wormhole are presented for two parameter sets: $a = 1.2, l = 4$ (left) and $a = 2, l = 4$ (right). In each subplot, two pairs of energy and angular momentum $(E, \ell)$ are shown with positive (solid lines) and negative (dashed lines) branches over the range $\rho \in [-10, 10]$. The results demonstrate that increasing the wormhole parameter $a$ amplifies the angular deviations of the trajectories. The positive and negative branches correspond to opposite angular directions, illustrating how the wormhole geometry influences photon paths. All calculations assume $\tilde{\kappa} = 0$ and an initial angular position $\phi(\rho_i) = 0$.}\label{fig:2}
\end{figure}

\begin{figure}[ht]
\centering
\includegraphics[scale=0.35]{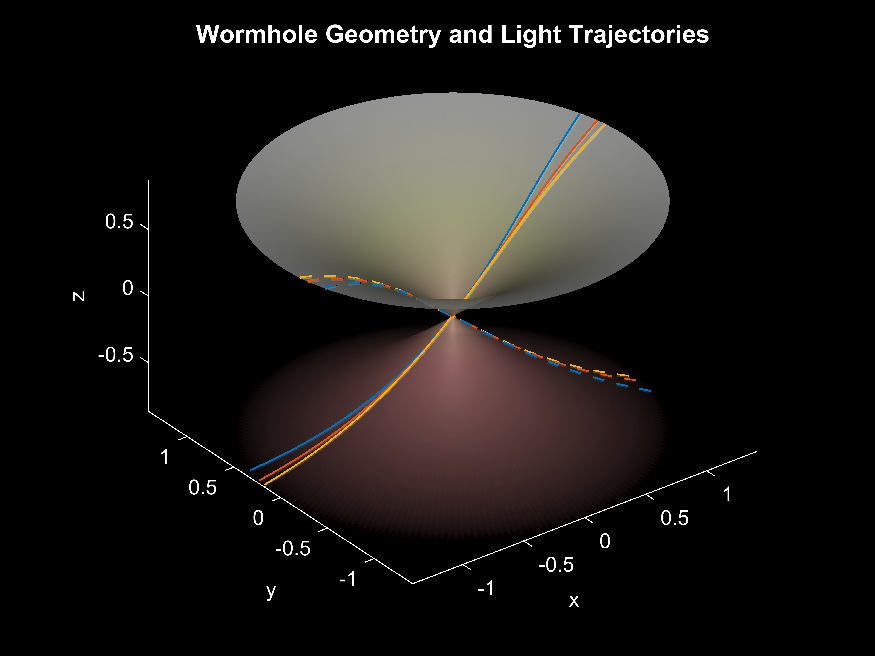}
\caption{\fontsize{7.7}{8.7}\selectfont Light trajectories threading a traversable wormhole. The wormhole geometry is defined by the embedding function \(\eta(\rho) = \ell \cosh\left(\frac{a}{\ell} \rho\right)\), with fixed parameters \(a = 1.2\) and \(\ell = 1.2\). Superimposed on the geometry are several null geodesics (light trajectories), plotted for three different energy values \(E = 2, 3, 4\), with angular momentum fixed at \(\ell = 1.2\). Each trajectory is color-coded: blue for \(E = 2\), orange for \(E = 3\), and yellow for \(E = 4\). Solid curves correspond to the positive trajectories, while dashed curves represent their mirrored negative counterparts. Here, we take $\rho$ in the range of $[-1,1]$ and $\tilde{\kappa}=0$.}
\label{fig:3D}
\end{figure}

Now, for arbitrary geodesics, we can obtain the following expression:
\begin{equation}
\dot{\phi} = \frac{1}{\eta(\rho) \, \sqrt{\frac{(\tilde{\kappa} + \mathcal{E}^2)}{\ell^2} \, \eta(\rho)^2 - 1}} \, \dot{\rho}.
\end{equation}
This equation defines the angular velocity $\dot{\phi}$ as a function of the radial motion, highlighting the coupling between the angular trajectory and the radial motion through the function $\eta(\rho)$ and the constants $\mathcal{E}$ and $\ell$. The effective potential impacts the angular motion by introducing a non-trivial dependence on the radial coordinate $\rho$. As a result, the angular trajectory $\phi(\rho)$ is derived as follows:
\begin{equation}
\phi(\rho) = \phi(\rho_i) \pm \int_{\rho_i}^{\rho_s} \frac{d\rho}{\eta(\rho) \sqrt{\frac{(\tilde{\kappa} + \mathcal{E}^2)}{\ell^2} \, \eta(\rho)^2 - 1}}.
\end{equation}
This integral governs the evolution of the angular coordinate \(\phi\) as a function of the radial position \(\rho\), with the integration limits \(\rho_i\) and \(\rho_s\) denoting the initial and final radial positions, respectively. Physically, this expression encapsulates how the angular displacement accumulates as the particle moves along its geodesic path under the influence of the background spacetime geometry. In particular, it provides insight into the rotational behavior of the particle as it traverses curved spacetime, accounting for the relationship between radial motion and angular motion. The exact analytical solution for \(\phi(\rho)\) is given by:
\begin{equation}
\begin{split}
&\phi(\rho) = \tilde{\phi}(\rho_i) \pm \frac{\ln \epsilon(\rho)}{a}, \quad \epsilon(\rho) = \frac{\tanh\left(\frac{a}{l} \rho\right)}{\sqrt{\tilde{a}}} + \sqrt{1 + \frac{\tanh^2\left(\frac{a}{l} \rho\right)}{\tilde{a}}}, \label{trajectories}
\end{split}
\end{equation}
where $\tilde{a} = \frac{l^2}{\ell^2} (\tilde{\kappa} + \mathcal{E}^2) - 1$ is a dimensionless parameter that encodes the combined effects of the underlying spacetime curvature, the conserved quantities and the type of test particle. The expression in Eq.~\eqref{trajectories} provides the exact angular trajectories \(\phi(\rho)\), and the corresponding geodesic paths are visualized in Figures~\ref{fig:2}--\ref{fig:3D}, highlighting how the conserved quantities, curvature, and geometry together shape the angular trajectories of rays.

\section{\mdseries{Wave optics}}\label{sec:4}
\begin{figure}[ht]
\centering
\includegraphics[scale=0.55]{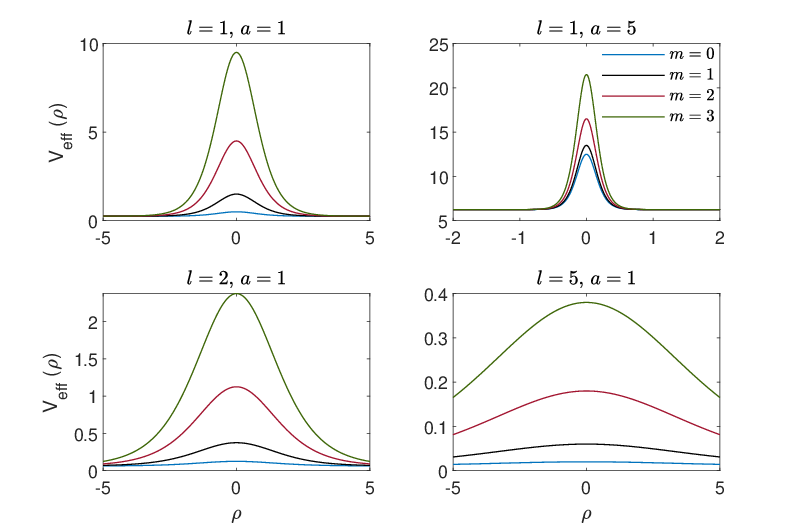}
\caption{\fontsize{7.7}{8.7}\selectfont Effective potential $V_{\text{eff}}(\rho) = \frac{a^2}{4l^2} + \frac{\frac{a^2}{4} + m^2}{l^2 \cosh^2\left(\frac{a}{l} \rho\right)}$ plotted as a function of the radial coordinate $\rho \in [-5,5]$ for different wormhole parameters and magnetic quantum numbers $m=0,1,2,3$. Each subplot displays the potential barrier profile for varying throat radius $l$ and wormhole parameter $a$. Increasing the magnetic quantum number $m$ raises the height and sharpness of the potential barrier. Larger $a$ values create taller and narrower barriers, reflecting stronger confinement effects near the throat, while increasing $l$ lowers and broadens the barrier, indicating weaker curvature influence. These results show how geometric parameters and magnetic quantum states govern the shape and strength of the effective potential barrier in wormhole geometries. Furthermore, narrower potential barriers are observed for smaller curvature radius (\( l / a \)).}
\label{fig:5}
\end{figure}
In the previous section, we derived the trajectories of light in the effective geometry of the optical wormhole. Now, we examine the corresponding propagating wave modes. Let us start by writing the Helmholtz equation in an effective geometry \cite{Rop2,N-2}:
\begin{equation}
    (\Delta_g + k^2)\Psi = 0,
\end{equation}
where \( \Delta_g \) is the Laplace-Beltrami operator and \( k \) is the wave number. For a generic metric \( ds^2 = g_{ij} dx^i dx^j \), the Laplace-Beltrami operator is given by \cite{Rop2,N-2}:
\begin{equation}
    \nabla_g \Psi = \frac{1}{\sqrt{g}} \partial_i \left( \sqrt{g} g^{ij} \partial_j \Psi \right),
\end{equation}
where \( g = |\text{det}(g_{ij})| \), \( i,j = \rho, \phi \). According to spatial part of Eq. \eqref{32}, the corresponding Helmholtz equation is obtained as follows:
\begin{equation}
\left[ \partial^2_{\rho}+\frac{a}{l}\,\tanh\left( \frac{a}{l}\rho \right)\partial_{\rho}+k^2-\frac{m^2}{l^2\,\cosh^2\left(\frac{a}{l}\rho\right)}\right]\,\psi(\rho) = 0,\label{WE}
\end{equation}
if \( \Psi(\rho,\phi) = \psi(\rho)\,e^{i\,m\,\phi} \), where \( m = 0, \pm 1, \pm 2, \dots \) due to the periodic boundary condition on \( \phi \). Let us now determine the effective potential for waves propagating in the optical wormhole background. To achieve this, we eliminate the first-order derivative term by defining:
\begin{equation}
\psi(\rho) = e^{-\frac{1}{2} \int \frac{a}{l} \tanh\left( \frac{a}{l} \rho \right) d\rho} \varphi(\rho).
\end{equation}
The exponential factor is carefully selected to ensure that, upon differentiation of \( \psi(\rho) \), the first-order derivative term cancels out seamlessly. This leads to the following simplified equation:
\begin{equation}
\ddot{\varphi}(\rho) + \left[k^2 - V_{\text{eff}}(\rho)\right]\varphi(\rho) = 0,\label{Sch-type}
\end{equation}
where
\begin{equation}
V_{\text{eff}}(\rho) = \frac{a^2}{4l^2} + \frac{\frac{a^2}{4} + m^2}{l^2 \cosh^2\left(\frac{a}{l} \rho\right)}.\label{eff-pot2}
\end{equation}
Eq. (\ref{Sch-type}) is a Schrödinger-like equation for the unknown function \( \varphi(\rho) \), with the effective potential given by Eq. (\ref{eff-pot2}). The profile of this effective potential is shown in Figure \ref{fig:5}, which exhibits a trend similar to that in Figure \ref{fig:1}. The effective potential (\ref{eff-pot2}) exhibits a repulsive nature, characterized by its peaked structure around \( \rho = 0 \) and a rapid decay toward a finite constant at asymptotically large \( |\rho| \). The term \( \frac{a^2}{4l^2} \) introduces a uniform frequency/energy shift, raising the baseline of the potential, while the second term, proportional to \( \cosh^{-2}\left(\frac{a}{l} \rho\right) \), dominates the spatial dependence. At \( \rho = 0 \), where \( \cosh^2\left(\frac{a}{l} \rho\right) = 1 \), the potential reaches its maximum value of
\begin{equation}
V_{\text{eff}}(0) = \frac{a^2/2 + m^2}{l^2} = V_0 > 0 \quad \text{(constant)}.\label{limit}
\end{equation}
Our analysis demonstrates that the medium can be described by a frequency- and space-dependent refractive index, \( n(\rho, \omega) \), derived from the generalized Helmholtz equation:
\[
\nabla^2 \varphi(\rho) + \frac{\omega^2}{c^2} n^2(\rho) \varphi(\rho) = 0,\quad k = \frac{\omega}{c}.
\]
The refractive index expression, based on the effective potential \( V_{\text{eff}}(\rho) \) in equation \eqref{eff-pot2}, is given by:
\begin{equation}
n(\rho, \omega) = \sqrt{1 - \frac{c^2}{\omega^2} \left( \frac{a^2}{4l^2} + \frac{\frac{a^2}{4} + m^2}{l^2 \cosh^2\left(\frac{a}{l} \rho\right)} \right)}.\label{RI}
\end{equation}
This expression accounts for the variation of the refractive index as a function of both the radial position \(\rho\) and the frequency \(\omega\), reflecting the influence of the effective potential \( V_{\text{eff}}(\rho) \). For wave propagation, the refractive index must remain real, which occurs when the argument inside the square root is non-negative. The critical frequency \(\omega_{\text{c}}(\rho)\), which makes $n=0$ and marks the transition between real and imaginary refractive indices, is given by:
\begin{equation}
\omega_{\text{c}}(\rho) = \frac{c}{\sqrt{V_{\text{eff}}(\rho)}}.\label{wc}
\end{equation}
For \(\omega < \omega_{\text{c}}(\rho)\), the refractive index becomes imaginary, leading to evanescent waves that decay exponentially with distance from the optical origin. For \(\omega > \omega_{\text{c}}(\rho)\), the refractive index is real, enabling wave propagation. At high frequencies (\(\omega \to \infty\)), the refractive index \eqref{RI} approaches 1, indicating that the medium's influence becomes negligible. As \(\rho \to \infty\), the refractive index tends to a non-zero value $0<n(\omega)<1$:
\[
n(\omega) \to \sqrt{1 - \frac{c^2}{\omega^2} \frac{a^2}{4l^2}},\quad \text{if} \quad \omega>\tilde{\omega}_c=\frac{c\,a}{2\,l}.
\]
Even in this limit, the propagation of waves with frequency $\omega < \tilde{\omega}_c$ is not permitted. This critical frequency, determined by the Gaussian curvature of the background, serves as the boundary between evanescent and propagating waves, with diffraction occurring for real refractive indices. Moreover, adjusting the background curvature can modify the variation of the refractive index, potentially altering the fundamental limits of diffraction. This may offer a means to control optical resolution, as demonstrated in recent studies \cite{Sugg-1}. Thus, in such an engineered medium, the refractive index can be tailored to control wave propagation and optical resolution.

\vspace{0.10cm}
\setlength{\parindent}{0pt}

Our exact solutions reveal that the curved optical wormhole geometry induces a spatially varying effective potential \( V_{\mathrm{eff}}(\rho) \), which modifies the local refractive index \( n(\rho,\omega) \) given by \eqref{RI} for waves of frequency \(\omega\). For wave propagation, the argument under the square root in \eqref{RI} must remain non-negative, leading to a local cutoff frequency \(\omega_c(\rho) = c / \sqrt{V_{\mathrm{eff}}(\rho)}\). Since \( V_{\mathrm{eff}}(\rho) \) peaks at the throat \(\rho=0\) with \( V_{\mathrm{eff}}(0) = (a^{2}/2 + m^{2}) / l^{2} \), the global minimum cutoff frequency is \(\omega_c^{\min} = c / \sqrt{V_{\mathrm{eff}}(0)}\). At large \(\rho\), the effective potential approaches \( V_{\mathrm{eff}}(\infty) = a^{2} / (4 l^{2}) \), corresponding to an asymptotic cutoff frequency \(\tilde{\omega}_c = 2 c / (a l)\). The spatial resolution is limited by the maximal local wavevector \( k n(\rho,\omega) \), where \( k = \omega / c \). In the high-numerical-aperture limit ($NA=n\sin \theta \to n$), where the numerical aperture (NA) approaches the refractive index, the diffraction-limited resolution follows the Abbe limit \cite{Abbe,Born}:
\[
\Delta x(\rho,\omega) \sim \frac{\pi}{k \, n(\rho,\omega)} = \frac{\lambda}{2 n(\rho,\omega)}.
\]
As \(\omega\) approaches \(\omega_c(\rho)\) from above, \( n(\rho,\omega) \to 0 \), causing \(\Delta x\) to diverge. This signals a loss of resolution and the onset of evanescent, non-propagating wave behavior. Therefore, waves with frequencies below $\omega_c^{\min}$ cannot propagate, establishing a fundamental cutoff for transmission. Increasing $a$ or $m$ raises $V_{\mathrm{eff}}(0)$, thereby increasing $\omega_c^{\min}$, which narrows the bandwidth and degrades resolution. Conversely, increasing $l$ lowers both quantities, allowing propagation of lower-frequency waves and improving resolution. Thus, the wormhole geometry and effective potential impose a spatially varying refractive index and a frequency-dependent diffraction limit. The throat cutoff frequency sets the minimal imaging frequency, with resolution deteriorating near this limit. Parameters $a$ and $l$ therefore provide tunable control over the optical bandwidth and imaging performance in condensed-matter analogs.

\section{\mdseries{Summary and discussions}}\label{sec:conc}

In this work, we conduct a detailed analysis of ray trajectories and wave optics in the optical geometry associated with the static BTZ black hole metric, as introduced in \cite{7}. This spacetime, cast in an ultra-static form, gives rise to an effective optical wormhole geometry, also termed a hyperbolic wormhole, characterized by a constant negative Gaussian curvature. Such geometries provide a compelling theoretical platform for investigating a wide range of physical phenomena, including the Hawking-Unruh effect, charge carrier dynamics, and decaying photonic modes within monolayer graphene systems and optical wormholes (see Figure \ref{fig:wormhole}).

\vspace{0.10cm}
\setlength{\parindent}{0pt}

Our analysis begins with the geometric optics regime, wherein light propagation is governed by null geodesics, corresponding to the limit $\tilde{\kappa} = 0$ in the equations of motion. The angular coordinate $\phi(\rho)$ as a function of the radial coordinate $\rho$ admits an explicit analytic form given by
\begin{equation*}
\begin{split}
\phi(\rho) = \tilde{\phi}(\rho_i) \pm \frac{1}{a} \ln \epsilon(\rho), \quad \epsilon(\rho) = \frac{\tanh\left(a/l\, \rho\right)}{\sqrt{\tilde{a}}} + \sqrt{1 + \frac{\tanh^2\left(a/l\, \rho\right)}{\tilde{a}}}.
\end{split} \label{eq:light-trajectories}
\end{equation*}
The function $\epsilon(\rho)$ encapsulates the radial variation of the optical trajectory and displays smooth monotonic behavior over the entire domain. In the vicinity of the wormhole throat, where $\rho \to 0$, the hyperbolic tangent approaches zero, and thus $\epsilon(\rho) \to 1$. Conversely, in the asymptotic region $\rho \to \infty$, the hyperbolic tangent approaches unity, and the function asymptotes to
\begin{equation*}
\epsilon(\rho) \to \tilde{C}, \quad \tilde{C} = \frac{1}{\sqrt{\tilde{a}}} + \sqrt{1+1/\tilde{a}}.
\end{equation*}
Accordingly, the angular coordinate $\phi(\rho)$ inherits this logarithmic structure. In the limit $\rho \to 0$, one recovers the initial angular value,
\begin{equation*}
\phi(\rho) \to \tilde{\phi}(\rho_i),
\end{equation*}
whereas in the asymptotic regime $\rho \to \infty$, the angular coordinate converges to
\begin{equation*}
\phi(\rho) \to \tilde{\phi}(\rho_i) \pm \frac{1}{a} \ln \tilde{C}.
\end{equation*}
These expressions describe the finite angular shift experienced by light rays during their propagation across the wormhole geometry. Without loss of generality, we may set $\tilde{\phi}(\rho_i) = 0$ to simplify the expressions, yielding $\phi(\rho) = \pm \frac{1}{a} \ln \epsilon(\rho)$ and a total deflection angle given by
\begin{equation*}
\Delta\phi = \phi(\infty) - \phi(0) \approx \pm \frac{1}{a} \ln \tilde{C}.
\end{equation*}
Figures~\ref{fig:2} and~\ref{fig:3D} provide a visual representation of the full trajectory structure under this parametrization.

\vspace{0.10cm}
\setlength{\parindent}{0pt}

Within this optical background, the radial motion of null geodesics is governed by an effective potential of the form
\begin{equation*}
V_{\text{eff}}(\rho) = \frac{\ell^2}{l^2 \cosh^2\left( \frac{a}{l} \rho \right)}, \label{eff-pot-conc}
\end{equation*}
which structurally resembles the well-known Pöschl-Teller potential, a prototypical exactly solvable model in quantum mechanics. This potential reaches its maximum at the wormhole throat $\rho=0$ and decays rapidly toward zero as $\rho$ increases, with the rate of decay modulated by the curvature scale and angular momentum parameter $\ell$. The profile of $V_{\text{eff}}(\rho)$, illustrated in Figure \ref{fig:1}, signifies a purely repulsive interaction, precluding the existence of physical and stable circular photon orbits.

\vspace{0.10cm}
\setlength{\parindent}{0pt}

We then extend our study to wave optics by solving the scalar wave equation in this background. Reformulating the problem in the form of a one-dimensional Schrödinger equation with an effective potential, we obtain
\begin{equation*}
V_{\text{eff}}(\rho) = \frac{a^2}{4l^2} + \frac{\frac{a^2}{4} + m^2}{l^2 \cosh^2\left(\frac{a}{l} \rho\right)}. \label{eff-pot2-conc}
\end{equation*}
Here, the ratio \( \frac{l}{a} \) sets the curvature scale of the optical geometry. Near the throat (\(\rho \to 0\)), the hyperbolic cosine function approximates unity, yielding the maximal potential value
\begin{equation*}
V_{\text{eff}}(0) = \frac{a^2}{4l^2} + \frac{a^2/4 + m^2}{l^2}.
\end{equation*}
In contrast, at large distances (\(\rho \to \infty\)), the second term decays exponentially due to the asymptotic form \(\cosh\left( \frac{a}{l} \rho \right) \approx \frac{1}{2} e^{\frac{a}{l} \rho}\), and the potential approaches a constant background value:
\begin{equation*}
V_{\text{eff}}(\rho) \to \frac{a^2}{4l^2}.
\end{equation*}
This asymptotic behavior reveals that the effective potential comprises a short-range barrier superimposed on a constant background. The curvature radius \( \frac{l}{a} \) governs both the spatial extent and the decay scale of the potential. At the optical origin, the peak value of the potential depends sensitively on the magnetic quantum number \( m \); however, at large radial distances, the contribution of \( m \) becomes negligible, although the constant term persists. For high \( m \), the central potential barrier becomes significantly pronounced, as depicted in Figure \ref{fig:5}.

\vspace{0.10cm}
\setlength{\parindent}{0pt}

We demonstrate that the optical response of this curved medium can be effectively described by a spatially and spectrally dependent refractive index \( n(\rho, \omega) \), derived from the generalized Helmholtz formalism. This index incorporates geometric effects via the effective potential:
\begin{equation*}
n(\rho, \omega) = \sqrt{1 - \frac{c^2}{\omega^2} V_{\text{eff}}(\rho)},
\end{equation*}
where \( V_{\text{eff}}(\rho) \) encodes the contribution of spacetime curvature. Wave propagation is allowed only if \( \omega > \omega_c(\rho) = \frac{c}{\sqrt{V_{\text{eff}}(\rho)}} \); otherwise, the solutions become non-propagating and exhibit exponential attenuation. In the asymptotic region \(\rho \to \infty\), the medium imposes a global cutoff frequency \( \tilde{\omega}_c = \frac{a\,c}{2\,l} \), independent of local spatial structure. In the high-frequency limit (\(\omega \gg \tilde{\omega}_c\)), the refractive index tends toward unity, rendering the medium effectively transparent. At intermediate frequencies, however, curvature-induced modulation of \( n(\rho, \omega) \) provides a mechanism to engineer wavefront shaping, control diffraction properties, and achieve spatial confinement. This curvature dependence suggests that spatial geometry itself may serve as an active medium for the control of wave propagation, with significant implications for metamaterial design, analog gravitational models, and subwavelength-scale photonic technologies \cite{semra-2025}. Furthermore, the wormhole’s curved geometry defines a local cutoff frequency, minimized at the throat, enforcing a fundamental low-frequency propagation threshold. Below this threshold, waves become evanescent and spatial resolution breaks down. Increasing curvature or angular momentum elevates this cutoff, restricting bandwidth and degrading resolution, whereas enlarging the throat reduces the cutoff, enhancing propagation and imaging fidelity. Hence, the geometric parameters $a$ and $l$ intrinsically control the refractive index landscape, diffraction limits, and spectral accessibility.

\vspace{0.10cm}
\setlength{\parindent}{0pt}

It is important to emphasize that graphene-based platforms exhibit exceptional geometric versatility, permitting configurations such as bending, rolling, or twisting to achieve desired curvature profiles. This controllability enables precise manipulation of effective curvature in monolayer graphene sheets \cite{12} and graphene-based optical wormholes \cite{R12}. Our results establish that the curvature radius directly modulates both the amplitude and spatial width of the effective background potential. Consequently, this control mechanism renders these structures highly promising for future applications in advanced photonic systems and engineered material platforms \cite{semra-2025,N-2,N-3}.

\section*{\small{CRediT authorship contribution statement}}

\textbf{Semra Gurtas Dogan}: Conceptualization, Methodology, Investigation, Writing – Review and Editing, Formal Analysis, Validation, Visualization, Project Administration.\\
\textbf{Abdullah Guvendi}: Conceptualization, Methodology, Investigation, Writing – Review and Editing, Formal Analysis, Validation, Visualization.\\
\textbf{Omar Mustafa}: Conceptualization, Methodology, Investigation, Writing – Review and Editing, Formal Analysis, Validation, Visualization.

\section*{\small{Data availability}}

No data were used to support this study

\section*{\small{Conflicts of interest statement}}

The authors have disclosed no conflicts of interest.

\section*{\small{Funding}}

This research has not received any funding.

\end{document}